\documentclass[usegraphicx]{mn2e}

\begin{document}

\title[Characterizing meteors]{Meteor light curves: the relevant parameters}

\author[N. Brosch et al.]{N. Brosch$^1${\thanks{noah@wise.tau.ac.il}}, Ravit Helled$^1$, D. Polishook$^1$,
E. Almoznino$^1$, N. David$^1$ \\ $^1$The Wise Observatory and the
School of Physics and Astronomy, the Raymond and  Beverly Sackler
Faculty of Exact Sciences, \\ Tel Aviv University, Tel Aviv 69978,
Israel}


\maketitle

\begin{abstract}

We investigate a uniform sample of 113 light curves of meteors
collected at the Wise Observatory in November 2002 during a
campaign to observe the Leonid meteor shower. We use previously
defined descriptors, such as the classical skewness parameter $F$
and a recently-defined pointedness variable $P$, along with a
number of other measurable or derived quantities, in order to
explore the parameter space in search of meaningful light curve
descriptors.

In comparison with previous publications, we make extensive use of
statistical techniques to reveal links among the various
parameters and to understand their relative importance. In
particular, we show that meteors with long-duration trails rise
slowly to their maximal brightness and also decay slowly from the
peak, while showing milder flaring than other meteors. Early
skewed meteors, with their peak brightness in the first half of
the light curve, show a fast rise to the peak. We show that the
duration of the luminous phase of the meteor is the most important
variable differentiating among the 2002 meteor trails.

The skewness parameter $F$, which is widely used in meteor light
curve analyses, appears only as the second or third in order of importance
in explaining the variance among the observed light curves, with the
most important parameter being related to the duration of the meteor
light-producing phase. We
suggest that the pointedness parameter $P$ could possibly be
useful in describing differences among meteor showers, perhaps by
being related to the different compositions of meteoroids, and also in
comparing observations to model light curves.

We compare the derived characteristics of the 2002 meteors with
model predictions and conclude that more work is required to
define a consistent set of measurable and derived light curve
parameters that would characterize the light production from
meteors, and suggest that meteor observers should consider publishing more
characterizing parameters from the light curves they collect.
Theorists describing the light production from meteors
should present their results in a form better compatible with
observations.

\end{abstract}

\begin{keywords}
meteors, light curves, statistical analysis
\end{keywords}

\section{Introduction}

In recent years, the study of meteors received significant input
primarily due to the campaigns to observe the Leonid meteor storms
(Jenniskens et al. 2000). The new studies introduced novel
observational methods as well as new analysis techniques. Among
those, the analysis of the light curves of visible meteors remains
one of the more widespread techniques.

The accepted picture regarding the light production by a meteor is
that collisions between the ablated meteor atoms and atmospheric
molecules and ions are responsible for this phenomenon. For small
meteoroids, the light production is proportional to the loss of
kinetic energy by the body. As the meteoroid velocity remains
almost constant throughout the luminous phase, it follows that the
light production tracks the instantaneous mass loss by ablation.

 A general review of meteor light curve analysis was presented by
 Hawkes et al. (2001). They discussed six improvements to traditional
 meteor light curve analysis in order to provide higher resolution
 and better information:
 use of generation III image intensifier technology, digital
 recording techniques, digital image processing algorithms which
 separate the even and odd video fields, whole-image background
 subtraction, analysis of pixel-by-pixel high resolution meteor
 light curves, and utilization of coincidence and correlation
 techniques. However, they did not introduce novel analysis
 methods of the higher quality data.

 The classical light curve (LC) produced by a solid, compact, and
non-fragmenting meteoroid should be smooth and exhibit its
maximum luminosity near the end of the trail (Cook 1954). This is
the combination of the exponential increase in air density
as the meteoroid penetrates into deeper atmospheric layers, and of the
reduction in the surface area presented by the
meteoroid to the airflow as the ablation proceeds. However, recent measurements
indicate that this picture of single-body ablation may not be
correct. In many cases, faint meteors were shown to produce
$\sim$symmetric LCs.

 A basic question concerning the behavior of the light produced by
 ablating meteors is therefore whether this process involves
 at all stages a single object, or whether during the production of light the
 meteoroid disintegrates in a rather large number of grains
 (the ``dustball'' model: Hawkes \& Jones 1975). This model assumes
 that the meteoroid is composed of numerous small grains with a
 high melting point temperature, held together by a low melting
 point glue.

 Fisher et al. (2000) followed Hawkes \& Jones (1975) and suggested
 that most meteoroids are collections
 of hundreds to thousands of fundamental minute grains, at least some of
 which are released prior to the onset of intensive ablation. One
 would expect these grains, unless extremely uniform in physical
 properties, to become aerodynamically separated during atmospheric entry, and
 therefore to produce a ``wake'', which is defined as some instantaneous meteor
 light production from an extended spatial region. Fisher et al. presented
 theoretical results for wake production as a function of grain mass
 distribution, height of separation, zenith angle and velocity.

 Koten \& Borovicka (2001) analyzed 234 meteor LCs, among which
 there were 110 Leonids from the 1998 and 1999 showers. One of
 their goals was the identification of relations between the LC
 shapes and other parameters. Among these, they included the
 leading and the trailing slopes of the LCs, defined as linear
 relations between the beginning of the LC and its maximum, and
 between the maximum and the terminal point of the LC.

 Bellot Rubio et al (2002) studied photographic light curves of
 relatively  bright meteors
 with magnitudes in the range from +2.5 to -5, collected by
 Jacchia et al. (1967) with Super Schmidt observations, in order to
 derive the average density of meteors and to test whether the single-body theory
 of meteor evolution fits the observations better than that of
 continuous disintegration. Velocities, decelerations and magnitudes were fitted
 simultaneously to synthetic light curves, and the ablation coefficient,
 the shape-density coefficient and the pre-atmospheric mass of each
 individual meteoroid were determined.
 Bellot Rubbio et al. could not confirm the large meteor
 density values determined from the quasi-continuous fragmentation models,
 essentially supporting the single-body ablation model.

 Babadzhanov (2002) reached opposite conclusions,
 supporting the continuous fragmentation model, from an analysis
 of 111 photographic light curves of meteors. Similar conclusions, that is a
 preference for continuous disintegration during re-entry of
 meteors, were reached by Jiang \& Hu (2001) from an analysis of
 high spatial resolution meteor light curves obtained during the
 1998 and 1999 Leonid showers.

 Light curves of Leonid meteors collected in recent years
 were analyzed by Murray et al. (2002). The meteors they
 concentrated on were fainter than those discussed by Bellot Rubbio
 et al. (2002), of 6-8
 mag, and the observations were performed with intensified CCD
 video cameras with fields of view from 16$^{\circ}\times$12$^{\circ}$
 to 40$^{\circ}\times$35$^{\circ}$. Murray et al. concentrated on
 the modified \emph{F} parameters and used these to distinguish among
 differences in overall light curve shapes. The \emph{F} parameter, to be
 explained below, ranged between 0.49 and 0.66 for individual annual
 showers (where \emph{F}=0.5 is defined as a symmetrical curve, \emph{F}$<$0.5
 an early skewed one, and \emph{F}$>$0.5 a late skewed curve).
 The findings indicate morphological differences between Leonid
 meteoroids observed in different
 years, thus originating from different ejection epochs.
 Murray et al. also noted the presence of very distinctive
(but not quantified) features among yearly events, with the 1998
Leonid light curves characterized by early skewed shapes
 while those from 1999 showed unusual ``flat topped'' curves. A similar
 conclusion was reached by Koten \& Borovicka (2001). In this, and
 in all the previously mentioned publications, the LCs were
 discussed almost exclusively in reference to the single quantitative
 parameter $F$, the skewness of the LC (Fleming et al. 1993).

 The ungluing of a complex of micro-meteoroids, with the
 subsequent formation of a classical trail, the epitome of the
 fragmentation model, was shown to fit the
 VHF radar observations of Geminid meteors (Campbell-Brown \&
 Jones 2003). The interesting finding of this study is the
 altitude at which the radius of the trail is zero, and which can
 be interpreted as the region where the 'ungluing' process begins:
 approximately 240-km for the Geminids. This fits the region where the high
 altitude radar echoes connected with meteor activity were detected by
 the Israeli L-band radar (Brosch et al. 2001).

 In addition, the light produced by many meteors was noted to be far from
 ``well-behaved'' and steady. Smith (1954) computed a simple model to
 account for the sudden brightening of meteor light curves. These
 flares cause the brightening of the meteor by 1-2 mag and, as
 explained by Smith, are
 the result of the release of a few thousand particles from the
 original single-body meteor.

 One explanation for this phenomenon,
 put forward by Kramer \& Gorbanev (1990), is that this flaring is produced
 by the shedding and spraying of a liquid film formed on the
 leading surface of the ablating meteor. They described a phenomenon
 by which the light intensity from the meteor is $\sim$constant,
 then shows a sudden drop (depression), after which the meteor flares in
 brightness. Therefore, meteor LCs may show both sudden brightening
 episodes, as well as sudden dimmings. Kramer \& Gorbanev remarked that
 meteors showing a depression in their LC have the brightness maximum earlier,
 on average, than meteors with no depression. In terms of the
 skewness parameter, such meteors would then be classified as ``early skewers''.

 It seems that in order to discuss statistical properties of
 meteors it is necessary to use descriptors that would reduce the
 amount of data characterizing a single LC while providing quantitative
 measures. The descriptors could
 be the symmetry parameter $F$ described above and used
 extensively, or the leading and trailing slopes as described by
 Murray et al. (1999) and by Koten \& Borovicka (2001). These could, in
 principle, refine the studies where tens to hundreds of LCs are
 collected and simultaneously analyzed, by reducing each LC to a
 small set of consistent numbers.

 The present paper presents an exploration of
 measurable parameters in the context of the examination of meteor
 LCs collected in Israel during the 2002 Leonid shower. A first
 attempt was already made in our contribution describing LCs of
 Leonids  2001, Geminids 2001, and Perseids 2002 (Brosch et al. 2002),
 where the pointedness parameter \emph{P} was introduced (see below).

The 2002 Leonid shower was analyzed by Arlt et al. (2002) with
 the following preliminary characteristics. The activity was due
 to two dust trails, one of cometary dust ejected seven
 revolutions ago in 1767 that produced a peak ZHR of 2510$\pm$60 and was
 seen in Asia and Europe, and the other by dust ejected four
 revolutions ago in 1866 that produced a ZHR of 2940$\pm$210 and was
 seen in the Americas. The first peak took place on
 November 19 at 04:10$\pm$1 UT with a full-width at half-maximum
 (FWHM) of 39$\pm$3 minutes. The second was in the same day at
 10:47$\pm$1 UT with a FWHM of 25$\pm$3 minutes. Our observations
 described below covered the period of the rise toward the twin
 peaks and a portion of the decay from the peak activity, but neither
 of the peaks themselves.

 Here we expand the discussion to include more parameters and to
 explore the internal correlations they might show for the case
 of the meteors observed in November 2002 during the Leonid shower
 by using classical statistical techniques. This
 approach is novel in the study of meteor LCs and may prove useful
 in uncovering hidden connections among the measured parameters, as
 many LCs are collected and uniformly analyzed in this manner. We
 describe our observations in section 2, the data reduction in
 section 3, the analysis and results in section 4, and conclude in
 section 5.

\section{Observations}

Starting in November 1998, the Wise Observatory (WO) is active in
the observation of meteors. From 2001 onward, these observations consist of
intensified video (ICCD) measurements, sometimes accompanied by
L-band phased-array radar observations. We do not discuss here the
radar observations (e.g., Brosch et al. 2001), but concentrate
exclusively on the analysis of the light curves (LCs) derived from
the intensified video observations.

The observations reported and analyzed here were collected during
five nights, from November 15-16 to November 19-20 2002, using
mobile meteor detection systems.  Each system is based on an ITT
Night Vision 18-mm, III-generation image intensifier (IT) with a
GaAs cathode, supplied by Collins Electro-Optics. The IT is
optically-coupled to an AstroVid 2000 CCD video camera operating
in the PAL TV standard (50 interlaced half-frames per second, or a
20 msec exposure for each half-frame). The IT is illuminated by a
50-mm f/0.95 Navitar lens and provides a final imaged field of
6$\times$8 degrees$^2$. The AstroVid 2000 sends the video
stream to a digital Hi8 video recorder equipped with a date/time
stamper and, in parallel, to a Matrox Meteor II frame grabber card
mounted in the docking bay of a Compaq Armada 850 computer. The
log of observations is given in Table 1 and shows the number of
meteors recorded by the observing stations during each individual
night as well as the total number of detected meteors.

\begin{table}
\caption{Meteor observations, by night and by camera}
\begin{tabular}{lccc}
\hline
Date & Cam1 & Cam2 & $\Sigma$ \\ \hline

15-16 Nov. 2002 & 54 & 34 & 88 \\

16-17 Nov. 2002 & 52 & 31 & 83 \\

17-18 Nov. 2002 & 67 & 18 & 85 \\

18-19 Nov. 2002 & 98 & 63 & 161 \\

19-20 Nov. 2002 & 53 & 35 & 88 \\ \hline

$\Sigma$ meteors   & 324 & 181 & 505 \\ \hline

\end{tabular}
\end{table}


The two cameras have somewhat different sensitivities. Camera 1
(Cam1) has a better IT with a clean and smooth field of view,
while camera 2 (Cam2) shows an off-center lower sensitivity spot
that covers $\sim$1\% of the cathode area. In general, camera 2
recorded only $\sim$60\% of the number of meteors recorded by
camera 1, except during the night of 17-18 November, when the
meteor brightness distribution was presumably different from that
during the other nights. Note that the preliminary analysis of the
2002 Leonid shower (Arlt et al. 2002) does not mention such a
possibility for the time preceding the first storm, though it does
mention a possibility of different population indices for bright
vs. faint meteors. Both cameras show a number of small dark spots;
these were identified, following a microscope inspection of the
AstroVid 2000 CCD chips, as minuscule solder droplets that block
one or a few adjacent pixels. The limiting magnitude for stellar
objects is $\sim$7.5 for Cam1 and 0.5 to 1 mag brighter for Cam2.

Prior to the beginning of observations in each night, the systems
are manually synchronized to Coordinated Universal Time (UTC) to
better than one second by using GPS receivers. The synchronization
is checked again at the end of the night, drifts are noted, and
the time stamp on the video recording is never found to deviate by
more than the accuracy of the original synchronization.

For the observations recorded here, the two cameras were operated
in non-tracking mode at the WO site and were roughly oriented
toward azimuth 050. Camera 1 was pointed to elevation $\sim$25 and
the Cam2 to elevation $\sim$19. This provided a slight overlap
between the two fields imaged by the cameras with the long
dimension aligned approximately with the horizon.

The observations were analyzed on-line using the MetRec program
(Molau 2001). The analysis allows for the on-the-fly detection of
meteors by comparing newly collected frames to an average of the
five previous frames. The meteors are tentatively identified with
one of the radiants active during the observing date, or as
sporadic meteors. The individual records can be identified at a
later date on the digital video tape and analyzed off-line, as
described below.

\section{Data reduction}

We selected for analysis 113 light curves, which are a fraction of
the 505 meteors observed. The LCs were selected to have the
beginning and end in the same field of one of the cameras.
Duplications (meteors showing up in both cameras) were not
rejected. A visual inspection of the list of meteors, which
includes their time-of-detection, revealed five cases when the
same object may have been detected by both cameras; these cases
were accepted as independent in the following analysis.

At this point, we checked to which shower does each meteor belong.
We relied mainly on the identifications performed by MetRec but,
in a few cases, we changed these so that a meteor identified by
MetRec in a certain way but showing similar angular velocity and
direction of travel as a group of meteors identified otherwise at
about the same time, would receive the identifier of the group.
The breakdown of recorded meteors by night and by shower is shown
in Table 2.

\begin{table}
\caption{Selected LCs by night and by shower}
\begin{tabular}{lcccccc}
\hline
Date  2002  & AMO   & LEO   & NTA   & STA   & SPO & $\Sigma$ by night \\
\hline

15-16 Nov.  & 1     & 8     & -     &  3    & 6     & 18 \\

16-17 Nov.  & -     & 20    &  1    & 1     & 9     & 31 \\

17-18 Nov.  & -     & 19    & -     & 1     & 2     & 22 \\

18-19 Nov.  & -     & 21    & 3     & -     & 6     & 30 \\

19-20 Nov.  & -     & 8     & 1     & 1     & 2     & 12 \\
\hline

$\Sigma$ by source & 1     & 76    & 5     & 6     & 25    & 113 \\
\hline

\end{tabular}
\end{table}

The source identifications in Table 2 are as follows:
AMO$\equiv\alpha$ Monocerotids, LEO$\equiv$Leonids,
NTA$\equiv$Northern Taurids, STA$\equiv$Southern Taurids, and
SPO$\equiv$sporadic meteors. Only for the Leonid and sporadic
sub-groups we collected sufficient LCs to warrant a separate
statistical analysis. Otherwise, the entire ensemble of meteor LCs
is treated together in what follows.

Below we present various analyses of this data set. Mostly, we do
not differentiate between meteors from different showers or the
sporadic ones, except when this is specifically mentioned in the
text. Note that out of 76 recorded Leonid meteors, 60 were
recorded by Cam1. The fraction of the sporadic meteors recorded by
Cam1 was 0.64.

For each selected event we extracted the individual frames from
the recorded video stream. We extracted a few frames before and
after the recorded meteor, in order to have a baseline image.
Following Fisher et al. (2000), we created a flat-field image by
which we divided the individual meteor images. This eliminates in
a major part the vignetting, the non-uniform illumination, and the
fixed patterns of the IT and CCD camera.

In order to compensate for short-duration changes in atmospheric
transparency we scaled all the images belonging to a single meteor
so that the few stars visible in the background would produce the
same number of net counts in each frame. We emphasize that we did
not scale the intensity so as to have fluxed readings. Finally,
the scaled brightness of the meteor on each frame was measured by
using a circular aperture wide enough to enclose the entire contribution
of the meteor.The local background was subtracted as a ring around
the meteor aperture, and the net brightness of the
meteor was plotted in instrumental magnitudes [-2.5$\times$log(total counts)]
against the frame
number in the collection of images. This is the final product that
was further analyzed, as follows.

\begin{figure}
\includegraphics[width=80mm, angle=-90]{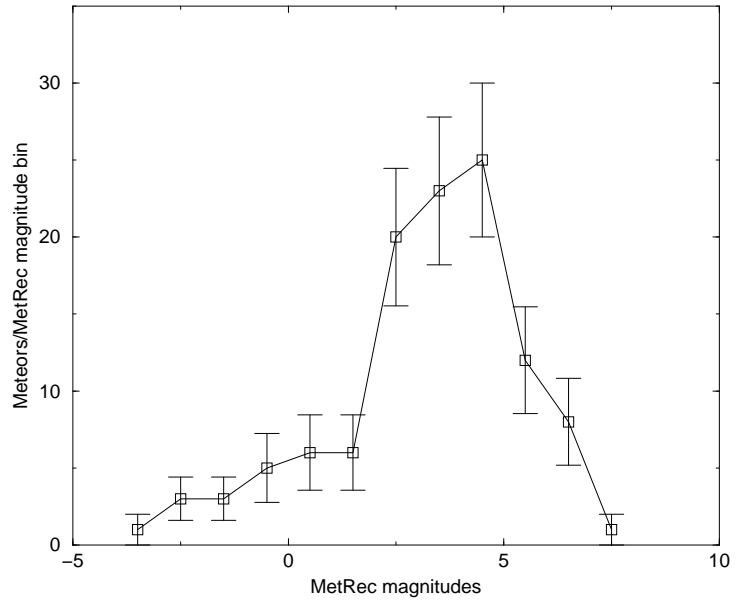}
\caption{Meteor magnitudes, as measured by MetRec. Note that these magnitudes
are different from those plotted in the LC of Fig. 2, and analyzed below. The error bars
assume Poisson statistics.}
\end{figure}
We note here that in a number of cases the meteor images were saturated.
In these cases we measured the brightness in the saturated portion of the image
by scaling from the non-saturated parts of the same image (usually the edges of the
saturated image). We scaled from the unsaturated part of the image to the
peak value by assuming the same image profile (transversal to the meteor train) as
for the unsaturated part of the image. We note also that this instrumental
magnitude is just a way of expressing the total counts detected by the cameras from the
meteor in logarithmic units and in each frame. This is different
from the classical meteor magnitude, as derived by MetRec from the initial
calibration at the beginning of the night.

Figure 1 shows the distribution of the meteor magnitudes as calculated by
MetRec with the statistical error in the number of meteors per bin
calculated assuming Poission statistics. Most meteors have magnitudes between
2 mag and 6 mag, as one could expect
from the experimental setup described above. Very few meteors (12 out of 113)
have negative magnitudes. The brightest meteor had a magnitude of -3.3 mag.
We note that these magnitudes, and the ``instrumental'' magnitudes mentioned
below, not being well-calibrated, are not suitable for a computation of the
meteoroid masses.

A smooth representation of each LC (sLC) was hand-drawn by NB in
order to eliminate the bumps and wiggles sometimes shown by the
meteors as well as flares, which could have influenced conclusions
regarding the general behavior of a LC. The sLC was measured by
RH. The consistency of the sLC fitting was tested by selecting ten
representative LCs and re-fitting them to a smooth representation
a few months after the first fit. The fit was done independently
by NB and by EA, and the goodness-of-fit was tested by calculating
the two shape parameters \emph{F} and \emph{P} (see below) and
comparing them between the two fits. We found that the internal
consistency of NB was $\sim$2--4\% and that between NB and EA was
$\sim$10\%. We estimate, therefore, that the shape parameters we
present and analyze, and the other variables included in the
analysis, carry experimental errors of this magnitude (10\%).

We defined a set of eight directly measurable parameters derived
from the original or from the sLC, as demonstrated in Figure 2 and
described below.

\subsection{Measured variables}
The parameters defined as ``measured variables'' are listed below:

\begin{enumerate}

\item $F$=symmetry, a parameter with no units. It is defined as
the ratio of the one-sided width of the LC at one mag below the
peak, to the total width at this level. Formally
\begin{equation}\label{}
    F=\frac{t_{B,1}-t_M}{t_{B,1}-t_{E,1}}
\end{equation}
where $t_{B,1}$ is the time near the beginning of the meteor trail
when its brightness is one magnitude fainter than the peak,
$t_{E,1}$ is the time near the trail end one magnitude below the
peak, and $t_M$ is the time of the peak brightness. Parameter $F$
is exactly that defined by Fleming et al. (1993) for a drop of one
magnitude below the peak.

\item $P$=pointedness, a parameter with no units. Defined as the
ratio of the sLC width at one mag below the (interpolated) peak
($\Delta t_{M1} \equiv T_{E, 1}-T_{B, 1}$) to the width at two mag
below the peak (${\Delta t_{M2}} \equiv T_{E, 2}-T_{B, 2}$):
\begin{equation}\label{}
    P=\frac{\Delta t_{M1}}{\Delta t_{M2}}
\end{equation}
The parameter $P$ is the equivalent of the \textsl{kurtosis}, the
fourth moment of a distribution, if we adopt the analogy of
likening $F$ to the \textsl{skewness} (the third moment of a distribution).

\item $M1$=magnitude difference between the first recorded point
of the sLC and the peak.

\item $M2$=magnitude difference between the last recorded point of
the sLC and the peak.

\item $JUMPS$=the largest frame-to-frame amplitude change in the
meteor brightness, in magnitudes, and limited to intensity
excursions higher than one magnitude. Measured from the individual
instrumental magnitudes in each frame, not from the smoothed
representation of the LC. \emph{JUMPS} measures the largest flare
or depression of the LC.

\item $DURATION$=time interval, in seconds, between the first and
the last detections of the meteor.

\item $D1$=time difference between the first point of the LC and
the peak (the duration of the leading part of the LC).

\item $D2$=time difference between the first point of the LC and
the peak (the duration of the trailing part of the LC). Obviously,
$DURATION=D1+D2$.

\end{enumerate}

Each observed meteor is characterized, therefore, not only by its
time of detection and its peak magnitude, but also by the set of
measurements of its LC.

Figure 2 demonstrates these measurable parameters using the LC of
a Leonid meteor detected on 18 November 2002 at 01:07:27UTC. This
is not a typical LC; it is the meteor trail with the longest
duration among the 113 LCs analyzed here, which implies that it is
described by the largest number of measurements. The LC is shown
as individual points; the smoother representation, shown here as
the solid line, is a plot obtained by smoothing the original LC
with a 12-point running average. Note that the beginning and the
end of the smooth representation used in this plot had to be
linearly extrapolated from the running mean smoothing. The meteor
magnitudes used for this plot are internal and refer to this specific field; they are
{\bf not} on the MetRec scale and repesent purely the logarithm
of the total counts produce by the meteor in each frame, scaled
to the brightness of the stars in the same frame. This magnitude allows us
to discuss relative changes of brightness of the same meteor.

\begin{figure}
\includegraphics[width=84mm, angle=-90]{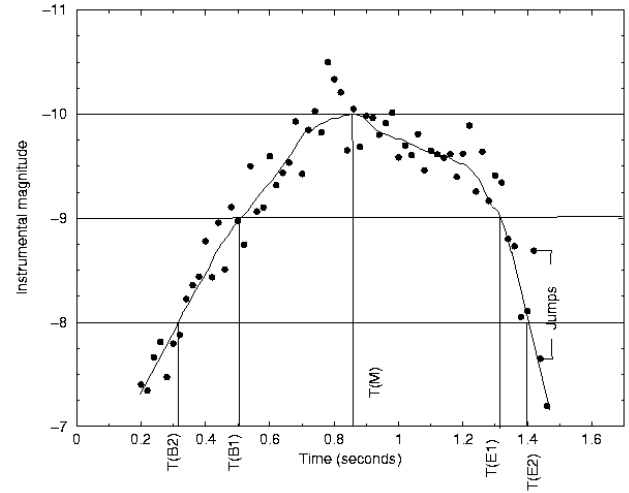}
\caption{Meteor light curve and measurable
parameters. We indicated the time (in frame numbers) from the
beginning of the meteor trail for T$_{B,2}$, T$_{B,1}$, T$_P$,
T$_{E,1}$, and T$_{E,2}$. We also marked the largest flare (or
sudden dimming) identified here as the variable $JUMPS$. The peak
magnitude of the sLC, as well as the values for one and two
magnitudes below the peak, are marked with horizontal lines. Note that
the instrumental magnitude here is different from the meteor magnitudes
plotted in Figure 1.}
\end{figure}

For completeness, we note that Campbell et al. (1999) attempted a
new definition of meteor LC parameters dealing with symmetry by
fitting a rotated parabola to the cubic root of the
meteor intensity. Although this apparently allows an objective fit
of a smooth curve to the LC that yields a measure of its symmetry,
it is hard to attach a physical meaning to the derived parameters.

\subsection{Derived parameters}
We define the $AMPLITUDE$ of the LC as the maximum of M1 and M2,
that is, the maximal magnitude difference between one of the end
points of the LC and its peak. \emph{AMPLITUDE} describes the
degree of brightening of the meteor relative to its faintest
recorded image.

We define the gradient of the leading part of the LC and that of
the trailing branch as the two slopes: $S1=M1/D1$ and $S2=M2/D2$.
The average and standard deviations of the slope values are
${S1}$=21.9$\pm$12.1 mag sec$^{-1}$ and ${S2}$=25.3$\pm$13.7
mag sec$^{-1}$. The two slopes describe the general speed of
brightening or of dimming of the meteor, irrespective of the
flaring phenomena.

We mentioned above that such LC slopes were mentioned by Murray et
al. (1999) and by Koten \& Borovicka (2001). In particular, Murray
et al. found a median leading slope (brightening) of 19.25 mag
sec$^{-1}$ and a median trailing slope (dimming) of --15.0 mag
sec$^{-1}$ (Camera M; their Table 2) for the 1998 Leonids observed
from the MAC-98 airborne platform.

As an example, we list here the parameters measured and derived
for the LC shown in Fig. 1: \emph{F}=0.51, \emph{P}=0.75,
\emph{M1}=2.48 mag., \emph{M2}=3.32 mag., \emph{JUMPS}=1.78 mag.,
\emph{DURATION}=1.28 sec., \emph{D1}=0.68 sec., \emph{D2}=0.60
sec., \emph{AMPLITUDE}=3.32 mag., \emph{S1}=3.64 mag sec$^{-1}$,
and \emph{S2}=5.55 mag sec$^{-1}$.

\section{Analysis and results}

We used SPSS (Statistical Package for Social Sciences) routines to
explore various statistical properties of the meteor LC
parameters. The analysis is as far as possible without
preconceptions and the parameters defined above are all treated
on an equal basis, without
searching a-priori for possible physical interpretations.
We first address the general distribution properties
of the variables. The shape parameters $F$ and $P$ have a mean and
standard deviation of 0.52$\pm$0.09 and 0.70$\pm$0.05
respectively, with $F$ being $\sim$normally distributed. The mean
observed duration of a meteor trail was 0.33$\pm$0.15 sec with a
maximal value of 1.28 sec. (for the meteor LC shown in Fig. 1).
Note that here and elsewhere in this paper there is a hidden bias
by our requirement that the meteor trails analyzed here be fully
included in the FOV of at least one camera. The average brightness
amplitude, defined as the maximal excursion from the beginning (or
the end) of a trail to its peak brightness, was 4.6$\pm$1.4 mag
with a maximal value of 8.08 mag. The maximal magnitude excursion
observed from frame to frame ($JUMPS$) was 3.93 mag and the
maximal values of the LC slopes \emph{S1} and \emph{S2} were 89
and 112 mag sec$^{-1}$, respectively.

Checking for differences among observing nights, we found that the
LCs from the night of 17-18 Nov. 2002 yielded meteor LCs with
smaller $AMPLITUDE$ than in other nights. While during the other
nights we observed the $AMPLITUDE$ variable range up to 7 and 8
mag., during this specific night the largest recorded value of
this parameter was 6 mag. Previously, in section 2, we already
showed that this specific night was different in that the meteor
magnitudes were biased toward fainter meteors than in other
nights. The reduction in \emph{AMPLITUDE} observed that night
might result from a combination of a fainter meteor population
with the lower detection threshold of Cam2 that would cause
fainter meteors not to be detected.

The time behavior of the measurable parameters during a single
night is also of interest. We found that $P$ showed a conspicuous
reduction in range with time during a night. While the full range
of $P$ before 01:00 LT every night was 0.62--0.85, the range
became restricted to 0.61--0.78 (with one outlier) after this
time. $F$ seems to show a similar behavior but one that is less
definite than $P$. $JUMPS$ was found to be more prominent after
01:00 LT, when we saw more values larger than 2.5 mag. $DURATION$
was more concentrated after this time ($\sim$0.3$\pm$0.1 sec) than
before this hour; this could be due to the higher fraction of
grazing trajectories before the Leonid radiant ascended, or to the
contribution of sporadic meteors. $AMPLITUDE$ had the opposite
behavior, namely a larger variety before 01:00 LT. The two LC
slopes, \emph{S1} and \emph{S2}, showed a similar behavior to that
of the \emph{AMPLITUDE} parameter, being steeper in the first half
of the night.


The distribution of shower identifications with the time during
the night is also relevant. Most Leonids were observed near dawn,
in the small hours of the morning, while the beginning of the
night saw mostly sporadic and Southern Taurid meteors. As
mentioned above, statistical properties of meteor populations are
limited to the Leonids and the sporadics. It is of interest to
compare the parameters of these two groups. Table 3 compares the
means of the various parameters by way of the 95\% confidence
interval within which the mean value resides.

\begin{table}
\caption{Mean (and standard deviation) parameters for LEO and SPO LCs}
{\begin{center}
\begin{tabular}{lcc}
\hline
   Parameter & LEO & SPO  \\ \hline
  $F$         &  .51(.09)   & .55(.08)  \\
  \emph{P}  & .70(.04) & .71(.05) \\
  $JUMPS$     &  2.2(.6)   & 1.9(.6)  \\
  $AMPLITUDE$ &  4.9(1.4)   & 3.9(1.4)  \\
  $DURATION$  & .31(.15)  &  .36(.12) \\
  \emph{S1} & 30(14) & 20(11) \\
  \emph{S2} & 34(17) & 29(17) \\ \hline

\end{tabular}
\end{center}}
\end{table}

Table 3 shows that the Leonid meteors have a more restricted range
of mean parameters, implying a more uniform set of characteristics
as a population than the sporadic meteors. There seems to be a
real difference in the leading and trailing slopes of the LCs; the
Leonids seem to show steeper slopes than the sporadic meteors. As
mentioned above, the sporadics are observed mainly during the
first part of the night while the Leonids appear in the second
half, with the rates increasing toward dawn. It is possible that
the relative velocity, compounded with the high incoming velocity
of the Leonids, may cause their faster ablation that would
translate into steeper LC slopes and a lack of meteors with long
\emph{DURATION}.

We searched for correlations among the primary variables (using
the bivariate correlation procedure) and present all the relevant
correlations in  Table 4. A few conclusions from the correlations that
are not immediately obvious are:
\begin{enumerate}
\item The brighter the meteor, from one of its LC end points to the maximum of the
LC, the larger the sudden brightness changes (flares) exhibited by its LC ($JUMPS$ vs.
$AMPLITUDE$).
\item The shorter the time span that the meteor was recorded,
 the more it tends to show violent change in brightness ($JUMPS$ vs. $DURATION$).
\item  The stronger a meteor LC flare is, the
 smaller is the $D1/D2$ ratio, thus the more tends the peak of the LC to
 be located near the beginning of the LC ($JUMPS$ vs. $D1/D2$). This is similar to the finding
 of Kramer \& Gorbanev (1990). Alternatively, we could express this
 as ``meteors with early skews of their LCs tend
 to flare stronger than other meteors''.
\item   It seems obvious that \emph{DURATION}
should correlate significantly with both the duration of the LC
from its start to the peak (\emph{D1}), and with the duration from
the peak to the end of the LC (\emph{D2}), but it is interesting
that the correlation with the leading
part of the LC is stronger than with the trailing side, past the maximum.
\item   The two parts of the LC,  $D1$ and $D2$,
correlate; the longer the rise to the peak, the longer
also the dimming from the peak brightness.
\item   The faster the
LC peak is reached, the faster also is the descent from the peak
($S1$ vs. $S2$).
\item The earlier the skew of an LC is, the steeper are the
leading and the trailing slopes of the LC ($S1$ and $S2$ vs. $F$).
\end{enumerate}

\begin{table*}
	    \begin{minipage}{160mm}
\begin{center}
\caption{Correlations among parameters}
{\small
\begin{tabular}{lccccccccccc}
Parameter & $F$ & $D1$ & $D2$ & $D1/D2$ & $JUMPS$ & $DURATION$ & $AMPLITUDE$ & $S1$ & $S2$ & $M1$ & $M2$ \\ \hline
$D1$ 	& .60** & - 	& - 	&-	& -	& -		& - 		& -	& -	& - & - \\
$D2$ & 	 -.67** & -.32**& -	& -	& -	& -		& -		& -	& -	& - & - \\
$D1/D2$ & .79**  & .74** & -.76**& -	& -	& -		& -		& -	& -	& - & - \\
$JUMPS$ & - 	& -	& .24*  & -.21  & -	& -		& -		& -	& -	& - & - \\
$DURATION$ & -	& -	& -	& -	& -.24* & -		& -		& -	& -	& - & - \\
$AMPLITUDE$ & -	& -	& -	& -	& .50** & -		& -		& -	& -	& - & - \\
$S1$ & -.41**	& -.23* & .34** & -.31**& .52**	& -.58**	& .55**		& -	& -	& - & - \\
$S2$ & .36**	& .28** & -.20* & .36** & .37** & -.53**	& .61**		& .49** & -	& - & - \\
$M1$ & -	& -	& .38** & -	& .21*  & .28**		& -		& .72** & .56** & .37** & - \\
$M2$ & -	& -	& -	& .26** & -	& .45**		& -		& .85** & .38** & .62** & .40** \\
\hline

\end{tabular}
}
\end{center}
Note{We use here the statistical notation of confidence: ** implies
99\% or higher confidence and * implies 95\% to 98\% confidence.}
	\end{minipage}
\end{table*}

We attempted to understand whether the LCs form a homogeneous
group, as defined by the measured and derived parameters, by
performing iteratively a K-means cluster analysis (KMCA). The KMCA
attempts to identify relatively homogeneous groups of LCs based on
selected characteristics, but requiring the prior specification of
the number of clusters the data should be split into.

For simplicity, we required a separation of the ensemble of 107
LCs that had all parameters available (six LCs had only part of
all the required parameters) into two clusters only. The results
indicate that cluster 1, with 76 of the 107 LCs, contains on
average meteors with small $JUMPS$ parameter, with $DURATION$ that
is $\sim$50\% longer that of the meteors in the second group, and
with shallower slope parameters, $\sim$half that of the LCs in the
second group. These LCs have, on average, eight more frames per
meteor (0.16 sec longer) than the LCs in cluster 2. The shape
parameters \emph{F} and \emph{P} do not appear to be different
among the two clusters. The number of members in each of the two clusters seems to
hint that this technique separates the LCs into LEO and SPO groups;
this is misleading, because the shower association was not one of the
variables used in the KMCA.

In order to identify the parameters that would be most promising
to differentiate among the LCs, we performed a series of factor
analyses using the Principal Component Analysis (PCA) routines of
SPSS. The PCA is a method of analysis that simplifies the way one
looks at a data set. Essentially, the PCA represents a rotation in
\emph{n}-dimensional event space (\emph{n} being the number of
variables that describe an event) in order to maximize the
variance of the data along the new axes. The PCA allows a
description of a data set using less variables than originally, by
choosing the new axes instead of the old. In our case, the events
are the 2002 LCs and the dimensions are the different variables,
measured or derived, that characterize each LC. As we chose to use
orthogonal (non-rotated) PCA, the new axes are linear combinations
of the variables, and are independent and orthogonal to each
other.

The PCAs included different sets of measured and derived
parameters, and the selection of the final factors was based on
the maximization of the explained variance among the LCs. The best
PCA, using only measured variables, was obtained with six
parameters that were grouped into three principal components, and
explained $\sim$90\% of the variance among the 113 LCs collected
in November 2002. Interestingly, the parameter $P$ that measures
the pointedness of the LC was not included in the final PCA. Table
5 presents the loadings of the variables making up the PCs, while
we show only those variables with loadings higher than 0.4. We are
aware of the danger of multi-collinearity by including variables
obviously related with each others (such as \emph{DURATION},
\emph{D1}, and \emph{D2}), but tests with the exclusion of one or
two of these while introducing the variable \emph{D1/D2} did not
show significant changes in the explained variance. Also, we are
using the PCA method only to indicate which variables could make
good characterizing agents for meteor LCs, and not in order  to
formulate a predictor formula to reproduce an LC.

\begin{table}
\caption{PCA including measured variables} {\begin{center}
\begin{tabular}{lccc}
\hline
   Variable & PC1 & PC2 & PC3 \\ \hline
  $F$         &  -    &   -   & 0.93 \\
  $AMPLITUDE$ &  -    & 0.74  & - \\
  $D1$        & 0.92  &  -    & - \\
  $D2$        & 0.84  &  -    & - \\
  $JUMPS$     & --0.42 & 0.74  & - \\
  $DURATION$  & 0.97  &  -    & - \\ \hline
Explained Variance (\%) & 46 & 23 & 21  \\ \hline

\end{tabular}
\end{center}}
\end{table}

The first Principal Component (PC1) contains information related
to the duration of the meteor LC: $DURATION, D1$, and $D2$, with
insignificant contributions from the other three parameters ($F,
AMPLITUDE$, and $JUMPS$). PC1 explains 46\% of the variance among
the LCs. PC2 explains another 23\% of the variance and contains
information about the brightness of the meteor through the
variables $AMPLITUDE$ and $JUMPS$. The third PC contains
information about the symmetry of the LC through the parameter
$F$, and adds 21\% to the explained variance. The first conclusion
from this analysis is that the parameter most commonly used to
explain the physics of meteor LCs, the symmetry parameter $F$,
seems not to be the most relevant in explaining the diversity of
meteors, at least when analyzing the 2002 LCs. The most relevant
component deals with the duration of the light-producing phase of
the meteor. The brightening of the meteor and its degree of
flaring are similar in relevance to the degree of symmetry of the
LC.

Another interesting finding is that the pointedness parameter $P$
did not enter the PCA at all. In fact, $P$ does not seem to
correlate with any other parameter. An analysis of variance
(ANOVA) shows that none of the other variables can be used to
predict $P$. These findings indicate that $P$ does not provide a
differentiating variable among the meteor light curves collected
in 2002, and that the parameter is not derivable from other
measurable LC quantities. We believe that the value of $P$ for
meteor characterization would be revealed in comparisons of
observed LCs to models, and in comparing meteor LCs from different
types of meteors, as show some of the finding reported below.

 The variable $JUMPS$ measures, as mentioned above, the highest
 brightness change from frame to frame, that is, within 40 msec,
 exhibited by the meteor. Table 4 shows that this variable is
 part of the PC dealing with the brightening of the meteor (at a marginal
level of significance), but
 also contributes part of the variance explained by PC1 that is
 connected to the duration of the meteor.

 We performed a second PCA, this time including the two derived slope
 variables $S1$ and $S2$ and all the variables used in the first PCA.
 We found that in this case we had to reject from the
 analysis not only $P$ but also $JUMPS$. On the other hand, the
importance of the symmetry parameter $F$ was now higher.
The resultant factor loadings are listed in Table 6. This resulted in a
 slight improvement, to $\sim$93\%, of the explained variance by
 increasing the contribution of PC1. It
 also reversed the importance, in terms of the percentage of the
 explained variance, of PC2 and PC3.

 \begin{table}
\caption{PCA with measured and derived variables} {\begin{center}
\begin{tabular}{lccc}
\hline
   Variable & PC1 & PC2 & PC3 \\ \hline
  $F$         &  -    &   0.95  & - \\
    $AMPLITUDE$ &  -    & -       & 0.81 \\
  $D1$        & 0.85  &  -      & - \\
  $D2$        & 0.83  &  -      & - \\
  $DURATION$  & 0.93  &  -      & - \\
  $S1$        & -0.78 & -       & - \\
  $S2$        & -0.74 & 0.53    & - \\  \hline
Explained Variance (\%) & 52 & 23 & 18 \\ \hline

\end{tabular}
\end{center}}
\end{table}

Table 6 shows that with the new variables introduced in the
analysis, the first PC still describes the time duration of the
visible meteor but now some additional influence comes from the
two slopes of the LC. As this variation shows up with negative
loadings, we conclude that the slopes influence the first PC in an
opposite sense to that of \emph{DURATION}, namely meteor LCs with longer
durations tend to have milder slopes. This, again, could be
connected with the ablation process: meteors showing shallower
slopes produce less light per unit time. As this light is an
expression of the ablation rate, assumed to be proportional to
$\frac{dm}{dt}$, meteors for which this fraction is smaller will
tend to last longer, i.e., produce light for a longer time
(assuming equal masses and velocities).

The second PC is now mostly influenced by the LC symmetry, with
significant influence from the slope of the trailing branch of the
LC through the $F$ parameter. The third PC is essentially the amplitude of the LC.
We can conclude this section by remarking that there are probably
a number of ways to describe light curves of meteors. It seems obvious that
merely using the symmestry parameter one does not realize the full
information content of the LCs. The final decision on what parameters are
important and what are not is a matter for further investigations.

 \section{Comparison with models}
The literature contains a small number of papers where model light
curves are presented, thus allowing in principle a comparison of theoretical
assumptions and calculations with experimentally-determined LCs.
Unfortunately, the model LCs do not contain the element of
sampling (threshold of detectability, for example), thus only a few of
the parameters used here (\emph{F} and \emph{P}) can be determined for these
model LCs. Moreover, the models deal with meteoroids of different masses; this
influence is disregarded in the comparison.

 Murray et al. (2000) produced a set of LC models following the
 dustball model of meteors proposed by Hawkes \& Jones (1975). We
 use these model LCs to derive parameters similar to those measured
 here to characterize the 2002 meteors. The parameter describing
 the different models of Murray et al. is the mass distribution
 index $\alpha$ that runs from 1.5 to 2.5. The LCs in Murray et al. are given
as magnitude vs. meteor altitude; given that the velocity of the
meteor hardly changes during the luminous phase, the altitude is
proportional to the time-of-flight variable used in our
characterization of the LCs.

We show in Table 7 the values of the two main parameters $F$ and
$P$ derived from the Murray et al. (2000) models as measured from
their plots. As mentioned above, we found for the 2002 meteors
${F}$=0.52$\pm$0.09 and ${P}$=0.70$\pm$0.05. Looking only at
Leonid and sporadic meteors, we find ${F}_{LEO}$=0.51$\pm$0.09
and ${P}_{LEO}$=0.70$\pm$0.04, and ${F}_{SPO}$=0.55$\pm$0.08
and ${P}_{SPO}$=0.71$\pm$0.05. The differences between the means
are not significant, given the size of the standard deviations,
thus the values do represent the meteor LCs
collected in 2002 irrespective of the meteor origin. Comparing
with models, this implies that $\alpha$ is either smaller than 1.5
or between 2 and 2.5, otherwise the measured values would not fit
the Murray et al. results. In particular, the pointedness
parameter \emph{P} measured here is very different from the values
calculated from the model LCs of Murray et al. We note that the
analysis of the visually-observed 2002 Leonids (Arlt  2003)
indicates a general background $\alpha$ (written as \emph{r} in Arlt et al.)
value of 1.9 with two maxima with population indices of 2.5$\pm$0.1 and
3.4$\pm$0.3.

\begin{table}
\caption{Parameters for model light curves (Murray et al. 2000)}
{\begin{center}
\begin{tabular}{ccc}
\hline
   $\alpha$ & $F$ & $P$ \\ \hline
     1.5 & 0.59 & 0.74 \\
     1.75 & 0.61 & 0.85 \\
     1.8 & 0.55 & 0.87 \\
     1.85 & 0.49 & 0.85 \\
     1.9 & 0.20 & 0.85 \\
     1.95 & 0.23 & 0.81 \\
     2.0 & 0.22 & 0.75 \\
     2.5 & 0.50 & 0.67  \\
 \hline

\end{tabular}
\end{center}}
\end{table}

Kuznetsov \& Novikov (2001) computed two generic models for
meteoroids ablating via the fast QCF (quasi-continuous
fragmentation) or via the slow QCF. Their LCs yield the following
shape parameters: \emph{F}=0.37 and \emph{P}=0.79 for the fast
QCF, and \emph{F}=0.40 and \emph{P}=0.73 for the slow QCF. Neither
of these models yields \emph{F} values close to those measured
here, though the \emph{P} for the fast-OCF seems to fit slightly
better than the other value.

Campbell \& Jones (2001) calculated light curves for meteoroids of
different masses. For the three masses they present in their Fig.
3, with mass indices $\alpha$=1.65, we measure from the plot:
\emph{F}=0.57 for all three LCs, and \emph{P}=0.79 for 10$^{-6}$
gr., 0.84 for 10$^{-7}$ gr., and 0.74 for 10$^{-8}$ gr. The lowest
mass bin with the smallest \emph{P} seems to fit best our
measurements for the 2002 meteors.

Lastly, Beech \& Murray (2003) produced models of ablating dustballs in connection
with the recent Leonid outbursts. We used their models to calculate \emph{F} and \emph{P}
parameters for the model light curves they presented in their Figures 6 and 7
and present the results in Table 8. The
two parameters were measured from the reproduced light curve plots. Note that their calculations
start integrating the meteor evolution at h(1)=120 km altitude. Therefore, their
light curves show no light production above 116 km, yielding artificially truncated
high-altitude behavior.

\begin{table}
\caption{Parameters for model light curves (Beech \& Murray 2003)}
{\begin{center}
\begin{tabular}{ccc}
\hline
   $\alpha$ & $F$ & $P$ \\ \hline
     1.0 & 0.64 & 0.79 \\
     1.65 & 0.50 & 0.92 \\
     1.73 & 0.46 & 0.92 \\
     1.83 & 0.45 & 0.85 \\
     2.0 & 0.40 & 0.83 \\
 \hline

\end{tabular}
\end{center}}
\end{table}

Note that Beech \& Murray (2003) calculated also the effect of the zenith angle on the
light curve. Their Figure 7 displays this effect for the $\alpha$=1.73 model with the
following resuts: Z=60$^{\circ}$; \emph{F}=0.48; Z=45$^{\circ}$: \emph{F}=0.45, \emph{P}=0.87; and
Z=0$^{\circ}$: \emph{F}=0.45, \emph{P}=0.70. As mentioned above, their LCs are cut off
at high altitude presumably by the evolution beginning at 120 km;
the effect is primarily causing high values for \emph{P} and the true
\emph{P} values cannot be easily recovered.


\section{Conclusions}
We presented an objective exploration of some measurable
parameters of meteor light curves based on a uniform collection of
LC from the 2002 Leonid shower. The LCs belong mostly to Leonid
and sporadic meteors and this allowed an inter-comparison of the
properties of these two groups. The analysis confirmed the reduced
importance of the symmetry parameter $F$ used in many previous LC
analyses, and showed that, at least for te meteor LCs studied here,
the most important parameters are the duration, the skewness, and
the brightness amplitude.

Comparisons of the distributions of measured parameters between
Leonid and sporadic meteors indicated that the LEO LCs are more
uniform than the SPO ones. The statistical analysis showed that
short-duration meteors have stronger sudden brightening and/or
dimming episodes (flares) than long-duration ones. The long
duration meteors have shallower general brightening and dimming
stages. Light curves with early skews, that reach their maximal
brightness before the middle of their luminous phase, tend to have
LCs with steep slopes.

A PCA showed that, at least for the meteors observed in 2002,
three principal components suffice to describe most of the
variation observed among the LCs. The components are related (in
order of importance) to the duration of a trail, its brightness
amplitude, and its amount of skewness. The ``flaring'' activity
presented by a meteor train seems to be related both to the
duration of the trail and to its brightness amplitude.

An attempt to compare the meteor LCs characterized here with
theoretical models was not very successful because of a lack of
published model light curves and of characterizing parameters,
such as those defined and tested here. However, the use of both
\emph{F} and \emph{P} in comparing the observations with the
models of Murray et al. (2000) indicated similar population
indices as estimated for the 2002 Leonids by Arlt et al (2002).
A comparison with the more recent models of Beech \& Murray (2003)
yielded a population index of $\alpha=1.6\pm0.1$, rather different from
that measured for the 2002 Leonids, but probably affected by the lack of
data describing the high altitude LC behaviour.

\section*{Acknowledgements}
We acknowledge support from the Israel Science Foundation to study
meteors in Israel. Discussions with Mr. Eran Ofek and Mr. Ilan
Manulis, and the continuing help of the Wise Observatory staff in
securing meteor observations, are appreciated. This paper has greatly benefited from the
remarks of an anonymous referee.


\end{document}